\magnification 1200
\baselineskip 18pt
\centerline{\bf ON THE ENERGY OF ROTATING GRAVITATIONAL WAVES$^*$}
\vskip1cm
\centerline{Bahram Mashhoon and James C. McClune}
\vskip.3cm
\centerline{Department of Physics and Astronomy}
\centerline{University of Missouri - Columbia}
\centerline{Columbia, MO 65211, USA}
\vskip1cm
\centerline{Enrique Chavez and Hernando Quevedo}
\vskip.3cm
\centerline{Instituto de Ciencias Nucleares}
\centerline{Universidad Nacional Aut\'onoma de M\'exico}
\centerline{A.P. 70 - 543} 
\centerline{04510 M\'exico D. F., M\'exico}
\vskip1cm
\centerline{\bf ABSTRACT}
\vskip 15pt
A class of solutions of the gravitational field equations
describing vacuum spacetimes outside rotating cylindrical
sources is presented. A subclass of these solutions corresponds
to the exterior gravitational fields of rotating cylindrical 
systems that emit gravitational radiation. The properties of these
rotating gravitational wave spacetimes are investigated.  
In particular, we discuss the energy density of these waves using
the gravitational stress-energy tensor.

\vskip.5cm\noindent
PACS numbers: 0420, 0430
\vskip1cm
$^*$ Work partially supported by DGAPA--UNAM, project IN105496

\vfill\eject\noindent
{\bf 1. INTRODUCTION}
\vskip 15pt
The main purpose of this paper is to present a class of cylindrically
symmetric vacuum solutions of the gravitational field equations representing
rotating gravitational waves and to study some of the physical properties
of such waves. The spacetimes under consideration here are not asymptotically
flat in general; therefore, the concepts of energy, momentum, and stress do
not make sense in the standard interpretation of general relativity.
Nevertheless, it is possible to introduce a local gravitoelectromagnetic
stress-energy tensor via a certain averaging procedure $[1]$.
This gravitational stress-energy tensor provides a natural 
physical interpretation of the Bel-Debever-Robinson tensor
that has been used frequently in numerical relativity [2]. 
In this work, we study the local gravitational stress-energy
 tensor for free rotating gravitational waves. 

Rotating cylindrically-symmetric gravitational waves were first discussed
in 1990 $[3, 4]$. The investigation of these solutions  was motivated by
Ardavan's discovery of the speed-of-light catastrophe $[5]$ 
and its implications
concerning gravitation $[6]$. Previous work is generalized in the present
paper and an extended class of rotating gravitational wave spacetimes is
analyzed. We find that the solution investigated previously 
$[1, 2]$ is
special, since it is the only member of the extended class studied  here
that represents the propagation of {\it free} rotating gravitational waves.

In this paper, we consider Ricci-flat spacetimes that are characterized by
a metric of the form

$$-ds^{2} = e^{2\gamma - 2\psi}(-dt^{2} + d\rho^{2}) + \mu^{2}e^{-2\psi}
  (\omega dt + d\phi )^{2} + e^{2\psi}dz^{2}, \eqno(1)$$
in cylindrical coordinates $(\rho,\ \phi,\ z)$. Here $\gamma,\mu,\psi$ and $\omega$ are
 functions of $t$ and $\rho$ only; moreover, the speed of light in
vacuum is set equal to unity except where indicated otherwise. The spacetime
represented by equation $(1)$ admits two commuting spacelike Killing vectors
$\partial_{z}$ and $\partial_{\phi}$. Though $\partial_{z}$ is 
hypersurface-orthogonal, $\partial_{\phi}$ is not and this fact implies
that the isometry group is not orthogonally transitive. Instead of the
variables $t$ and $\rho$, it is convenient to express the gravitational field
equations in terms of retarded and advanced times $u=t-\rho$ and $v=t+\rho$,
respectively. The field equations then take the form

$$(\mu\psi_{v})_{u} + (\mu\psi_{u})_{v} = 0, \eqno(2)$$

$$\mu_{uv}-{l^{2} \over 8}\mu^{-3}e^{2\gamma} = 0, \eqno(3)$$

$$\omega_{v}-\omega_{u} = l\mu^{-3}e^{2\gamma}, \eqno(4)$$

$$\gamma_{u}={1 \over 2\mu_{u}}(\mu_{uu} + 2\mu\psi^{2}_{u}), \eqno(5)$$

$$\gamma_{v}={1 \over 2\mu_{v}}(\mu_{vv} + 2\mu\psi^{2}_{v}), \eqno(6)$$
where $\psi_{u} = \partial\psi /\partial u$, etc. Here $l$ is a constant
length characteristic of the rotation of the system. Using equation $(3)$,
it is possible to eliminate $\gamma$ from equations $(5)$ and $(6)$; then,
we obtain the following equations

$$2\mu^{2}\psi^{2}_{u} = 3\mu^{2}_{u} - \mu\mu_{uu} + \mu\mu_{u}\mu_{uuv}
  (\mu_{uv})^{-1}, \eqno(7)$$

$$2\mu^{2}\psi^{2}_{v} = 3\mu^{2}_{v} - \mu\mu_{vv} + \mu\mu_{v}\mu_{uvv}
  (\mu_{uv})^{-1}. \eqno(8)$$
The integrability condition for this system --- i.e. $\psi_{uv}=\psi_{vu}$ --- results in a nonlinear fourth order partial differential
 equation for 
$\mu$. Alternatively, one could obtain the same equation for $\mu$ by
combining equations $(2),\ (7),$ and $(8)$, which shows the consistency of
the field equations $(2) - (6)$. It is important to notice that $\mu$ cannot
be a function of $u$ or $v$ alone, since this possibility would be
inconsistent with equation $(3)$. The partial differential equation of
fourth order for $\mu$ is, however, identically satisfied if $\mu$ is a
separable function, i.e. $\mu=\alpha (u)\beta (v)$; this leads, in fact, to
the rotating waves discussed earlier $[3, 4]$. Here we wish to study the
{\it general} solution of the field equations.

In section $2$, we discuss the general solution of the field equations. We
find that there are two possible classes of solutions: The first class
corresponds to the stationary exterior field of a rotating cylindrical 
source, while the second class appears to represent a mixed situation
involving rotating gravitational waves. In fact, a subclass of the latter
solutions describes the exterior fields of certain rotating sources that
emit gravitational radiation; these solutions approach the special rotating
wave solution $[3, 4]$ far from their sources. Indeed, the only solution
of the second class corresponding to a pure gravitational wave spacetime is
the special solution $[3, 4]$ that is further discussed in section $3$. In section $4$, some aspects of the energy and momentum of the special rotating
gravitational waves are discussed using the gravitoelectromagnetic
stress-energy tensor developed in a recent work $[1]$. The appendices contain
some of the detailed calculations.
\vskip1cm

{\bf
\noindent 2. SOLUTION OF THE FIELD EQUATIONS
}
\vskip 15pt
To solve the field equations $(2)-(6)$, let us introduce the functions
$U, V$, and $W$ by

$$U = \mu_{u}\gamma_{u} - {1 \over 2}\mu_{uu}, \eqno(9)$$

$$V = \mu_{v}\gamma_{v} - {1 \over 2}\mu_{vv}, \eqno(10)$$

$$W = \mu \gamma_{uv} + {3 \over 2}\mu_{uv}, \eqno(11)$$
and rewrite equations $(5)$ and $(6)$ as

$$\mu\psi_{u} = \epsilon \ (\mu U)^{1/2},\qquad
 \mu\psi_{v} = \hat{\epsilon} \ (\mu V)^{1/2}, \eqno(12)$$
where the symbols $\epsilon$ and $\hat{\epsilon}$ represent either $+1$ or $-1$ (i.e.,
$\epsilon^{2}=\hat{\epsilon}^{2}=1)$. Using equation $(3)$, it is straightforward to show
that

$$\mu U_{v} = \mu_{u}W,\qquad  \mu V_{u} = \mu_{v}W. \eqno(13)$$

\noindent
Let us now combine relations $(12)$ and $(13)$ in order to satisfy equation
$(2)$; the result is

$$\epsilon  \ U^{-1/2} (\mu_{v}U+\mu_{u}W)+\hat{\epsilon} \ V^{-1/2} (\mu_{u}V+\mu_{v}W) =
  0. \eqno(14)$$
This equation can be written as

$$U(\mu_{u}V+\mu_{v}W)^{2} = V(\mu_{v}U + \mu_{u}W)^{2}, \eqno(15)$$
which holds if either $W^{2} = UV$ or $\mu_{u}^{2} \ V = \mu_{v}^{2} \ U$; that is,
equation $(15)$ can be factorized into a fourth order equation and a third 
order equation, respectively.

Let us first consider the fourth order equation $W^{2} = UV$. It follows from
this relation and equation $(12)$ that $W = \pm\mu\psi_{u}\psi_{v}$. This
expression for $W$ along with $U = \mu\psi_{u}^{2}$ and $V = \mu\psi_{v}^{2}$
can be inserted into equation $(13)$ to obtain relations that when combined
with equation $(2)$ in the form

$$2\mu\psi_{uv} = -(\mu_{u}\psi_{v} + \mu_{v}\psi_{u}) \eqno(16)$$
imply

$$\psi_{u}\psi_{v} = 0. \eqno(17)$$
If $\psi$ is a function of either $u$ or $v$ alone, then equation $(2)$
implies that $\psi$ must be a constant. The spacetime represented by a
constant $\psi$ turns out to be a common special case of the solutions 
discussed in this section and is {\it flat}.

Let us next consider $\mu_{u}^{2} \ V = \mu_{v}^{2} \ U$. It follows from
equation $(3)$ that

$$\gamma_{u} = {3 \over 2} {\mu_{u} \over \mu} + {1 \over 2}
  {\mu_{uvu} \over \mu_{uv}}, \eqno(18)$$

$$\gamma_{v} = {3 \over 2} {\mu_{v} \over \mu} + {1 \over 2}
  {\mu_{uvv} \over \mu_{uv}}, \eqno(19)$$
which together with equations $(9)$ and $(10)$ imply that $\mu_{u}^{2} \ V =
\mu_{v}^{2} \ U$ is essentially equivalent to

$${\mu_{uvv} \over \mu_{v}} - {\mu_{vv}\mu_{uv} \over \mu_{v}^{2}} =
  {\mu_{uvu} \over \mu_{u}} - {\mu_{uu}\mu_{uv} \over \mu_{u}^{2}}.\eqno(20)$$
Equation $(20)$ can be written as

$$(\ln\mu_{v})_{uv} = (\ln\mu_{u})_{vu}. \eqno(21)$$
Let $f(u)$ and $g(v)$ be arbitrary functions of their arguments
 and consider
the transformation $(u, v) \to (x, y)$ such that

$$x = f(u) + g(v), \eqno(22)$$

$$y = f(u) - g(v), \eqno (23)$$
$f_{u}\not= 0$, and $g_{v} \not= 0$; then, the general solution of equation
$(21)$ is that $\mu$ should not depend on $y$, i.e.

$$\mu = \mu(x). \eqno(24)$$

Let us now proceed to the calculation of $\psi$. To this end, let 
us consider a function ${\cal L} (x)$ given by

$${\cal{L}}^{2}(x) = {1 \over 2} \left(3\mu{'}^{2} - \mu\mu{''} + 
\mu {\mu{'}\mu{'''}
 \over \mu{''}}\right), \eqno(25)$$
where a prime indicates differentiation with respect to $x$.
Equations $(7)$ and $(8)$ can now be written as

$$\mu^{2}\psi^{2}_{u} = f^{2}_{u} \ {\cal L}^{2}, \qquad \mu^{2}\psi^{2}_{v} =
  g^{2}_{v} \ {\cal L}^{2}. \eqno(26)$$
Furthermore,

$$\psi_{u} = {\left( {\partial\psi \over \partial x} + {\partial\psi \over
  \partial y} \right)} f_{u}, \eqno(27)$$

$$\psi_{v} = {\left( {\partial\psi \over \partial x} - {\partial\psi \over
  \partial y} \right)} g_{v}. \eqno(28)$$
Combining equations $(26)-(28)$, we find that either $\psi$ is purely a
function of $x$ given by $\mu^{2} (d\psi /dx)^{2} = {\cal L}^{2}(x)$ or $\psi$
is purely a function of $y$ given by $\mu^{2}(d\psi /dy)^{2} = {\cal L}^{2}
(x)$. These cases will now be discussed in turn.
\vskip.5cm

{\it Case (i)}: $\psi = \psi(x)$

It follows from equation $(2)$ that in this case

$${d\psi \over dx} = {C \over \mu (x)}, \eqno(29)$$
where $C$ is an integration constant. Thus $\mu (x)$ is determined by
${\cal L}^{2}(x) = C^{2}$. To solve this differential equation, let $X=\mu{'}$
and note that

$${\cal L}^{2} = {1 \over 2} \, X^{2} \left(\mu^{2} {dX \over d\mu}\right)^{-1} 
\ {d \over d\mu} \left(\mu^{3}{dX \over d\mu}\right) .\eqno(30)$$

\noindent
Moreover, let $ S = \mu \ dX/d\mu$; then, ${\cal L}^{2}(x) = C^{2}$ can be
written as

$${d S \over dX} = 2 \left( {C^{2} \over X^{2}} -1 \right) , \eqno(31)$$
which can be integrated to give $ S = -2(D+X+C^{2}/X)$. Here $D$ is an
integration constant. It follows from this result that $\mu (x)$ can be found
implicitly from $d\mu = X(\mu) dx$, where

$$\mu^{2} = \exp\left( - \int {XdX \over X^{2}+DX+C^{2}}\right) .\eqno(32)$$
We will show that for $C=0,\ \psi$ is constant and this solution
turns out to be flat. If $C\not= 0$, however, we have a new solution.

The spacetime metric in this case can be written in a form that depends only
on $x$. That is, it is possible to show --- by a transformation of the metric to normal form --- that the general solution in Case ($i$) has an 
extra timelike or spacelike Killing vector field $\partial_{y}$. To this end, let
us consider the coordinate transformation $(t,\rho ,\phi, z) \to (T, R,\Phi, Z)$, 
where $f(u)=(T+R)/2,\ g(v)=(-T+R)/2,\ \phi = \Phi + H(T,R)$, and $z=Z$. 
In this transformation $H(T, R)$ is a solution of the partial differential
equation

$$- (F+G) {\partial H \over \partial T} + (F-G) {\partial H \over \partial R}
  + {2 \over l} (F+G) {d\mu \over dR} = 0, \eqno(33)$$
where $F=f^{-1}_{u},\ G=g^{-1}_{v}, R=x$, and $T=y$. To prove our assertion,
we will show that under this coordinate transformation the spacetime metric
in Case ($i$) takes the form

$$-ds^{2} = P(R) (-dT^{2} + dR^{2}) + \mu^{2}(R)e^{-2\psi (R)} [\Omega (R)
  dT+d\Phi ]^{2}+ e^{2\psi (R)} dZ^{2}, \eqno(34)$$
which is clearly invariant under a translation in $T$ thus implying the 
existence of a Killing vector field $\partial_{T}$. If $P(R) > 0$, then this
metric could represent the {\it stationary} exterior field of a rotating cylindrical 
configuration. It follows from a comparison of the metric forms $(1)$ and
$(34)$ that

$$P(R) = -{2 \over l^{2}} \mu^{3} \ {d^{2}\mu \over dR^{2}} \ e^{-2\psi (R)},
  \eqno(35)$$

$${\partial H \over \partial T} = - {1 \over 4}\omega (F-G)+\Omega (R), \eqno (36)$$

$${\partial H \over \partial R} = - {1 \over 4}\omega (F+G). \eqno(37)$$
Equations $(36)$ and $(37)$ can be used in equation $(33)$ to show that
$\Omega (R) = 2l^{-1}d\mu /dR$. It remains to show that the integrability
condition for equations $(36)$ and $(37)$, i.e. \hfill\break
$\partial^{2} H/\partial R\partial T = \partial^{2}H/\partial T\partial  R$, is 
satisfied. It turns out that
this relation is indeed true, since it is equivalent to the field equation
$(4)$ for $\omega$.

Finally, let us consider the case $C=0$ or, equivalently, $\psi =$ constant.
It follows from equation $(31)$ that in Case ($i$)

$${d^{2}\mu \over dR^{2}} = -{2 \over \mu} \left[\left( {d\mu \over dR}
  \right) ^{2} + D{d\mu \over dR} + C^{2}\right] ,\eqno(38)$$
and one can show explicitly 
that the metric form $(34)$ is flat once $C=0$. For
$C\not= 0$, the general solution of Case ($i$) is not flat; however, its 
physical properties will not be further discussed in this work, 
which is devoted to rotating {\it gravitational waves}.

\vskip.5cm
{\it Case (ii)}: $\psi = \psi (y)$

It follows from $(d\psi /dy)^{2} = {\cal L}^{2}(x)/\mu^{2}$ that $\psi$ must
be a linear function of $y$, since the left hand side of this equation is
purely a function of $y$ while the right hand side is purely
a function of $x$; therefore, each side must be constant. Thus 
$\psi = ay+b$, where $a$ and $b$ are constants, and 
${\cal L}^{2} = a^{2}\mu^{2}$.
It turns out that equation $(3)$ is identically satisfied in this case. Using
equation $(30)$, the differential equation for $\mu$ can be expressed in
terms of $X=\mu{'}$ as

$$X^{2} {d \over d\mu} \left( \mu^{3}{dX \over d\mu}\right) = 2a^{2} \mu^{4}
{dX \over d\mu}.\eqno(39)$$
If $a=0$, $\psi$ is constant and we recover the same flat spacetime solution
as in Case ($i$) with $C=0$. Therefore, let $a\not= 0$ and consider a new
``radial'' coordinate $r$ given by $r=(a^{2}\mu^{2})^{-1}$; then,

$$r^{2}X^{2} {d^{2}X \over dr^{2}} + {dX \over dr} = 0. \eqno(40)$$
The solution $X =$ constant is unacceptable, since it implies that $\gamma =
-\infty$. However, there is another exact solution that is given by

$$X=\pm \left( {3\over 2} \ r \right)^{-1/2}, \eqno(41)$$
which turns out to correspond to the special rotating gravitational waves
$[3,4]$ that are the subject of the next section. It is possible to transform
equation $(40)$ to an autonomous form; to this end, let us define $A$ and $B$
such that $A = 2/(3rX^{2})$ and $B=-4X_{r}/(3X^{3})$, where $X_{r}=dX/dr$.
It can then be shown that equation (40) is equivalent to

$${dB \over dA} = {3 \over 2} {B(B-A^{2}) \over A(B-A)}, \eqno(42)$$
where $(A,B) = (1,1)$ represents the special solution $(41)$.
This special solution is an isolated singularity of the nonlinear autonomous
system $(42)$ and the behavior of characteristics near this point indicates
that $(1,1)$ is a saddle point.

Let us note here certain general features of $X(r)$, which is the solution of
equation $(40)$. If $X(r)$ is a solution, then so is $-X(r)$. Moreover, 
equation $(40)$ can be written as $(X^{2}_{r})_{r} = -2X^{2}_{r}/(r^{2}X^{2})$,
which indicates that for $X\not=$ constant the absolute magnitude of the
slope of $X(r)$ monotonically decreases as $r$ increases. If $X(r)$ has a zero
at $r_{0}\not=0$, then the behavior of $X(r)$ near $r_{0}$ is given by

$$X(r) = \pm\sqrt{2\over r_{0}} \left( {r\over r_{0}} -1 \right) ^{1/2}
  \left[ 1+{1\over 2} \left( {r\over r_{0}} -1 \right) - {3 \over 76}
  \left( {r\over r_{0}} -1 \right) ^{2} + \dots \right] ,\eqno(43)$$
for $r\ge r_{0}$. These results are illustrated in figure $1$.

Once a solution $X(r)$ of equation $(40)$ is given, one can find a solution
of the field equations in Case ($ii$). It turns out that in general such a solution is not
a pure gravitational wave. The only exception is the special
solution $(41)$. To demonstrate this, let us first consider a transformation
of the spacetime metric $(1)$ to the normal form appropriate for Case $(ii$).
The first step in this transformation will be exactly the same as that given 
in Case ($i$) and equation $(33)$, i.e. 
$(t, \rho , \phi , z) \to (T, R, \Phi , Z)$.
Next, we change the radial coordinate to $r$ and let 
$(T, R, \Phi , Z) \to (\hat{t}, r, \Phi , \hat{Z})$, 
where $\hat{t} = 2aT$ and $\hat{Z} = la^{2} \exp (2b) Z$. 
In terms of the new system of coordinates, the spacetime metric may be 
expressed --- up to an overall constant factor --- as

$$-ds^{2} = e^{-\hat{t}} \ {X_{r} \over X} \ \left( -X^{2} d\hat{t}^{2} + 
{dr^{2} \over   r^{3}} \right) + {e^{-\hat{t}} \over 
\hat{l}^{2}r} (\hat{l} X d\hat{t} + d\Phi)^{2}
  + e^{\hat{t}} d\hat{Z}^{2}, \eqno(44)$$
where $\hat{l} = (la)^{-1}$ is a constant and $X(r)$ is a solution of equation
$(40)$. There are four curvature invariants for this Ricci-flat spacetime and
it is possible to show that two of these are identically zero. The other two
also vanish for the special solution $(41)$; therefore, this special case
corresponds to {\it free} gravitational waves $[7]$. Specifically, let

$$I_{1}=R_{\mu\nu\rho\sigma}R^{\mu\nu\rho\sigma} -
i R_{\mu\nu\rho \sigma}R^{*\mu\nu\rho\sigma}, \eqno(45)$$

$$I_{2}=R_{\mu\nu\rho\sigma}R^{\rho\sigma\alpha\beta}
        R_{\alpha\beta} ^{\ \ \, \mu\nu} + i  
R_{\mu\nu\rho\sigma}R^{\rho\sigma\alpha\beta}
  R^{*\ \ \mu\nu}_{\alpha\beta}, \eqno(46)$$
be the complex invariants of spacetime. We find that for the 
metric form $(44)$ these are real and are given by

$$I_{1} = -e^{2\hat t} (r X^{4}X^{2}_{r})^{-1} (1 - 3rX^{2} - 2r^{2}XX_{r} -r^{3}
  X^{3}X_{r} + r^{4}X^{2}X^{2}_{r}), \eqno(47)$$

$$I_{2} = - {3 \over 4} e^{3\hat t}(rX^{5}X^{3}_{r})^{-1} (1-2 rX^{2} - 2r^{2}XX_{r}
  + 2r^{3}X^{3}X_{r} + r^{4}X^{2}X^{2}_{r}). \eqno(48)$$
It is important to note that $I_{1}$ and $I_{2}$ vanish for the special solution
$X=\pm (3r/2)^{-1/2}$, 
indicating that it describes the propagation of free gravitational waves.
Moreover, all other solutions --- which do not describe
{\it free} gravitational waves --- are singular for $\hat{t}\to \infty$, since 
$I_{1}$ and $I_{2}$ both diverge in the infinite future.
Thus all such solutions evolve to states that are ultimately singular. The
singular nature of the special solution $(41)$ has been discussed previously
$[3,4]$.

It is clear from equation $(40)$ and figure 1 that as $r \to 0,\ X(r) \to
\pm (3 r/2)^{-1/2}$ for a subclass of the spacetimes under consideration here.
Such solutions correspond to the exterior field of a rotating and radiating 
cylindrical source such that very far from the symmetry axis the metric nearly
describes free rotating gravitational waves given by the unique special
solution $(41)$. Indeed, inspection of the metric form $(44)$ reveals that
the circumference of a spacelike circle about the symmetry axis and normal to it
is proportional to $r^{-1/2}$ at a given time $\hat{t}$; therefore, as $r \to 0$
the metric form $(44)$ describes the asymptotic region very far from the axis.
Moreover, it is simple to check that $I_{1} \to 0$ and $I_{2} \to 0$ as
$r \to 0$ for the solution given in figure $1$ with $X_{r} (r=1) = -1$.

Finally, let us note that the unique special solution corresponding to free
gravitational waves $(41)$ can be written in terms of $\mu$ and $x$ as
$d\mu /dx = \pm(2/3)^{1/2}a\mu$, which can be easily solved to show that $\mu$
depends exponentially upon $x=f(u) + g(v)$. This means simply that $\mu$ can be
written as $\mu = \alpha (u)\beta (v)$, where $\alpha$ and $\beta$ are
arbitrary functions. The next section is  devoted to a discussion of the 
properties of this solution beyond what is already known from previous studies
$[3, 4]$.

\vskip1cm
{\bf
\noindent 3. FREE ROTATING GRAVITATIONAL WAVES
}
\vskip 15pt
Free nonrotating cylindrical gravitational wave solutions of Eintein's
equations were first discussed by Beck $[8]$. These solutions have been the
subject of many subsequent investigations $[7, 9, 10]$. The 
solutions considered in this section can be interpreted in terms of
simple cylindrical waves that rotate $[4]$.

The special solution (41) for free rotating gravitational waves is given by

$$\mu = \alpha (u)\beta (v), \eqno(49)$$ 
$$\psi = \sqrt{3 \over 2} \ln{\alpha\over\beta}, \eqno(50)$$

$$\gamma = {1\over 2} \ln \left[ {8\over l^{2}}(\alpha\beta )^{3}\alpha_{u}
  \beta_{v} \right] , \eqno(51)$$

$$\omega_{v} - \omega_{u} = {8\over l} \alpha_{u}\beta_{v}. \eqno(52)$$

\noindent
It turns out that in this case equation $(2)$ is equivalent to the scalar
wave equation for $\psi$ in the background geometry given by equation $(1)$;
therefore, the function $\psi$ --- which is a mixture of ingoing and outgoing
waves according to equation $(50)$ --- has the interpretation of the scalar
potential for the free rotating gravitational waves. The solution $(49)-(52)$
cannot be thought of as a collision between outgoing and ingoing gravitational
waves, since the field equations do not admit solutions for which $\mu = 
\mu (u)$ or $\mu = \mu (v)$. That is, there is no purely outgoing solution
just as there is no purely ingoing solution.

The spacetime given by equations $(49)-(52)$ is singular. In fact, 
the analysis of the corresponding curvature indicates that  moving 
singular cylinders appear whenever $\alpha , \beta , \alpha_{u}$, or
$\beta_{v}$ vanishes. It is interesting to consider the nature of the
symmetry axis for rotating waves, since in these solutions the axis does not in
general satisfy the condition of elementary flatness. If for an infinitesimal
spacelike circle around the axis of symmetry the ratio of circumference to
radius goes to $2\pi$ as the radius goes to zero, the condition of elementary
flatness is satisfied for the axis under consideration. In our case, this
means that $\mu^{2}/(e^{2\gamma}\rho^{2}) \to 1$ as $\rho \to 0$. For simple
cylindrical waves (i.e., Beck's solution) we have $\mu = \rho$ and hence the 
condition of elementary flatness is that $\gamma \to 0$ as $\rho \to 0$. In
general, Beck's fields can be divided into two classes: the Einstein-Rosen
waves $[9]$ and the Bonnor-Weber-Wheeler waves $[10]$. In the former class,
the axis does not satisfy the condition of elementary flatness; in fact, the
axis is not regular either. It is a singularity of spacetime and is therefore
interpreted as the source of the cylindrical waves which are otherwise free of
singularities. In the latter case, the axis does satisfy the condition of
elementary flatness and is, moreover, regular. The waves presumably originate
at infinity: Incoming waves implode on the axis and then move out to infinity
with no singularities in the finite regions of spacetime. It is therefore 
clear that {\it no} caustic cylinders appear regardless of the nature of the
axis. They do appear, however, when the waves {\it rotate}. We are therefore led
to regard the appearance of moving singular cylinders in our solution as being
due to the rotation of the waves. It is important to emphasize the absence
of a direct causal connection between the violation of elementary flatness
at the axis and the presence of singular cylinders: the axis is static while
the singularity is in motion, and the condition of elementary flatness involves
the gravitational potentials ($g_{\mu\nu})$ while the singularity of the field
has to do with the spacetime curvature. In fact, it is possible to find
instances of exact solutions for which the axis is elementary flat but singular
$[11]$.

In spacetime regions between the extrema of $\alpha$ and $\beta$, the solution
$(49)-(52)$  can be reduced to a normal form in which $\alpha$ and $\beta$ are
linear functions of their arguments $[3]$. With further elementary coordinate
transformations, the normal form can be reduced to two special solutions which
are of interest: ($a)\ \alpha = \sigma u, \beta = \sigma v$ and 
$(b)\ \alpha = 
\sigma u, \beta = -\sigma v$. Here $\sigma = \pm (l_{0})^{-1/2}$, where 
$l_{0}$ is a constant length whose introduction is necessary on dimensional
grounds. The metrics for these cases are, respectively,

$$-ds^{2} = {8 \over l_{0}^{4}l^{2}}(uv)^{3} 
\left( {u\over v}\right)^{-\sqrt 6}
  (-dt^{2} + d\rho^{2})+ {1\over {l_{0}^{2}}}(uv)^{2} 
\left( {u\over v}\right)^{-\sqrt 6}
  \left( {8\over l_{0}l} \rho dt + d\phi\right)^{2} $$
$$+ \left( {u\over v}\right)  ^{\sqrt 6} dz^{2} \eqno(53)$$
and

$$-ds^{2} = -{8 \over l_{0}^{4}l^{2}} (-uv)^{3} \left( -{u \over v}\right)
  ^{-\sqrt 6} (-dt^{2}+d\rho^{2}) + {1\over {l^{2}_{0}}}(uv)^{2} \left( -{u \over v}\right)
  ^{-\sqrt 6} \left( -{8 \over l_{0}l} \rho dt + d\phi \right)^{2}$$
$$+  \left(- {u\over v}\right)  ^{\sqrt 6} dz^{2}. \eqno(54)$$
To ensure that the spacetime metric is real, case $(a)$ must be limited to
$(u > 0, v > 0)$ or $(u < 0, v < 0)$. Similarly, case $(b)$ must be limited to
$(u > 0, v < 0)$ or $(u < 0, v > 0)$. In either case the hypersurfaces
$u = 0$ and $v = 0$ are curvature singularities. Since the laws of physics
break down very close to these surfaces, it appears that the consideration of
boundary conditions across such surfaces would be without physical 
significance. In case $(b)$, $\rho$ is a temporal coordinate and $t$ is a 
spatial coordinate.
The transformation $(t \to \rho,\ \rho \to t)$, or equivalently 
$(u \to -u,\ v \to v)$, brings the metric to the form

$$-ds^{2} = {8 \over l_{0}^{4}l^{2}}(uv)^{3} \left( {u\over v}\right)
  ^{-\sqrt 6} (-dt^{2} + d\rho^{2}) + {1\over {l^{2}_{0}}}(uv)^{2} \left( {u\over v}\right)
  ^{-\sqrt 6} \left( -{8\over l_{0}l} td\rho + d\phi\right)^{2} $$
$$+  \left( {u\over v}\right)  ^{\sqrt 6} dz^{2}, \eqno(55)$$
which reduces to case $(a)$ with a further coordinate transformation
$\phi \to \phi + 8t\rho /(l_{0}l)$. Therefore, only the normal form in
case $(a)$ will be considered in the rest of this paper.

The geodesic equation for free rotating gravitational wave spacetimes is
discussed in appendix A.
\vskip1cm

\noindent
{\bf 4. THE GRAVITATIONAL STRESS-ENERGY TENSOR
}
\vskip 15pt
In a recent paper $[1]$, a local gravitoelectromagnetic stress-energy tensor
$T_{\mu v}$ has been defined for Ricci-flat spacetimes via a certain averaging
procedure in a Fermi frame along the path of a geodesic observer. That is,
$$T_{\mu\nu} = {L^{2} \over 12\pi} T_{\mu\nu\rho\sigma} \lambda^{\rho}_{(0)}
  \lambda^{\sigma}_{(0)}$$
$$= {L^{2} \over 12\pi} T_{\mu\nu(0)(0)} \ , \eqno(56)$$
where $L$ is a length-scale characteristic of the field under consideration,
$T_{\mu\nu\rho\sigma}$ is the Bel-Debever-Robinson tensor $[7, 12]$ defined by

$$T_{\mu\nu\rho\sigma} = {1 \over 2}(R_{\mu\xi\rho\zeta}
R_{\nu\ \sigma}^{\ \xi\ \zeta} + R_{\mu\xi\sigma\zeta}
R_{\nu\ \rho}^{\ \xi\ \zeta} ) - {1\over 16}  g_{\mu\nu}g_{\rho\sigma}R_{\alpha\beta\gamma\delta}
R^{\alpha\beta\gamma  \delta}, \eqno(57)$$
and $\lambda^{\mu}_{(0)} = dx^{\mu}/d\tau$ is the vector tangent to the timelike path of the observer. 
The gravitational stress-energy tensor is expected to provide {\it approximate} measures of the average gravitoelectromagnetic energy, momentum, and stress in the neighborhood of the observer.
There does not seem to be any direct connection between
$T_{\mu\nu}$ and the Landau-Lifshitz pseudotensor; this point is further
discussed in appendix B.

It would be interesting to study the energy of rotating gravitational waves.
However, the spacetime metric for the rotating waves $(44)$ is given 
explicitly
only for the case of the exact solution $(41)$ of the differential equation
$(40)$. This unique solution is approached asymptotically by the rotating wave solutions very far from the symmetry axis.  For the sake of simplicity, we therefore restrict 
attention to the special solution $(41)$ that, according to the 
results presented in section 3, 
corresponds to the {\it free} rotating waves.

To determine the energy density of this radiation field measured by a geodesic
observer, it is first necessary to consider a solution of the geodesic
equation discussed in appendix A. Let us therefore choose a ``radial'' geodesic
of the free rotating waves given by
$$t = {l_{0} \over 2}(\zeta^{1/p} + \zeta^{1/q}), \eqno(58)$$

$$\rho = -{l_{0} \over 2} (\zeta^{1/p} - \zeta^{1/q}), \eqno(59)$$

$$\phi = {\sqrt 5 \over 2} \left( \zeta^{2/p} - \zeta^{2/q} - 
\sqrt{3 \over 2}\   \zeta^{4/5} \right) + \phi_{0}, \eqno(60)$$

$$z = z_{0}, \eqno(61)$$
where we have used equations $(A11) - (A14)$ of appendix A together with the
metric in normal form (53), i.e. $\alpha (u) = u/l_{0}^{1/2}$ and $\beta (v) =
v/l_{0}^{1/2}$, and we have fixed the constants in these equations such that
$\tau_{1} = \tau_{2} = l_{0}$ and 
$$2c_{1} = l_{0}^{1-p/2}, \qquad 2c_{2} = l_{0} ^{1-q/2}\ .$$

\noindent
Note that the scaling parameter $l_{0}$ is therefore related to the intrinsic
rotation parameter $l$ via $2l_{0} = \sqrt{5} l$. Moreover, $\phi_{0}$ and
$z_{0}$ are constants and $\zeta$ is defined by

$$\zeta = 1-\tau /l_{0}. \eqno(62)$$
The observer under consideration here starts at the symmetry axis $\rho = 0$, which is regular
at $\tau = 0$, and returns to it at $\tau = l_{0}$ when it is singular. The
trajectory of this observer $\rho (\phi)$ is depicted in figure 2.

Imagine now an orthonormal tetrad frame $\lambda^{\mu}_{(\alpha)}$ that is
parallel propagated along this timelike worldline. It is simple to work out
explicitly two axes of the tetrad: 
the time axis and the spatial axis parallel 
to the symmetry axis. These are given by $\lambda^{\mu}_{(0)} = dx^{\mu}/
d\tau$ using equations $(58)-(61)$ and

$$\lambda^{\mu}_{(3)} = \zeta^{-3/5} \ \delta^{\mu}_{3}, \eqno(63)$$
respectively. It follows from the projection of $T_{\mu\nu\rho\sigma}$ given
by equation $(57)$ on these axes that along the geodesic the gravitational
radiation energy density is given by

$$T_{(0)(0)} = {36 \over 625\pi} {L^{2} \over (l_{0} - \tau)^{4}},\eqno(64)$$
while the energy flux along the symmetry axis vanishes, $T_{(0)(3)}= 0$, as
expected. There is, however, radiation pressure along the $z$-axis and is 
given by

$$T_{(3)(3)}= {3 \over 125\pi} {L^{2} \over (l_{0}-\tau)^{4}}. \eqno(65)$$
The energy density and the pressure measured by the geodesic observer both
diverge at the singularity $\tau = l_{0}$. 
Note that $T_{(3)(3)}/T_{(0)(0)} = 5/12$, which is less than
 unity as would be expected for the ratio of pressure to density. 
The characteristic
length-scale associated with rotating gravitational waves is $l$; therefore, 
the constant length $L$ in expressions $(64)$ and $(65)$ could be chosen to be 
simply proportional to $l = 2l_{0}/ \sqrt 5$.
\vskip1cm
{\bf
\noindent APPENDIX A: GEODESICS IN ROTATING WAVE SPACETIMES
}
\vskip 15pt
Starting with the metric form given by equation $(1)$, let

$${\cal L}_g = -{1 \over 2}\left ({ds \over d\lambda}\right )^{2} 
= {1 \over 2} 
[e^{2\gamma -2\psi} (-\dot{t}^{2} + \dot{\rho}^{2}) + \mu^{2}e^{-2\psi} 
(\omega\dot{t} + \dot{\phi})^{2} + e^{2\psi} \dot{z}^{2}], \eqno (A1)$$
where $\dot{t} = dt/d\lambda$, etc. Since the Lagrangian $(A1)$ does not 
depend upon $\phi$ and $z$, there are two constants of the motion $p_{\phi}$ 
and $p_{z}$ given by

$$p_{\phi} = {\partial{\cal L}_g \over \partial\dot{\phi}} = \mu^{2}
  e^{-2\psi} (\omega\dot{t} + \dot{\phi})\ , \eqno (A2)$$

$$p_{z} = {\partial{\cal L}_g \over \partial\dot{z}} = e^{2\psi} \dot{z}.
  \eqno (A3)$$
Moreover, we have

$${d \over d\lambda} (\omega p_{\phi} -e^{2\gamma -2\psi} \dot{t}) =
  {\partial {\cal L}_g \over \partial t}, \eqno (A4)$$
and

$${d \over d\lambda} (e^{2\gamma -2\psi} \dot{\rho}) = 
{\partial{\cal L}_g \over \partial\rho}. \eqno (A5)$$ 

Let us now focus attention on ``radial" geodesics such that the 
momenta
associated with azimuthal and vertical motions vanish. It is then convenient 
to write the equations of motion in terms of radiation coordinates $u$ and 
$v$.  We find that equations $(A2) - (A5)$ reduce to

$${d \over d\lambda} \left ( e^{2\gamma -2\psi} {du \over d\lambda}\right ) =
  (e^{2\gamma -2\psi})_v \ {du \over d\lambda} {dv \over d\lambda} \eqno
  (A6)$$
and

$${d \over d\lambda} \left ( e^{2\gamma -2\psi} {dv \over d\lambda}\right ) =
  (e^{2\gamma -2\psi})_u \ {du \over d\lambda} {dv \over d\lambda}, \eqno(A7)$$
which imply that 
$$e^{2\gamma -2\psi} \ {du \over d\lambda} {dv \over d\lambda}$$ 
is a constant along the path. This constant vanishes for a null 
geodesic, so that ``radial" null geodesics correspond to $u=t-\rho =$ constant 
or $v = t+\rho =$ constant, where $\lambda$ is the affine parameter along the 
path.

We are mainly interested in timelike ``radial" geodesics, hence we set
$\lambda = \tau$, where $\tau$ is the proper time along the path. Then, 
with $\tilde U = du/d\tau$ and $\tilde V= dv/d\tau$, we have 
$\tilde U\tilde V = e^{2\psi -2\gamma}$. It 
follows from $(A6)$ and $(A7)$ that 
$\tilde U_v \tilde V ^2  = \tilde V _u \tilde U ^2$. 
The general solution is given by

$${du \over d\tau } = {1 \over 2} \left ( {\partial {\cal S} \over \partial u}
\right )^{-1},  \eqno (A8)$$

$${dv \over d\tau} = {1 \over 2} \left ( {\partial {\cal S} \over \partial v}\right )
^{-1},  \eqno (A9)$$
where ${\cal S}(u, v)$ is any solution of the differential 
equation

$$4 \ {\partial{\cal S}\over \partial u} {\partial{\cal S}\over \partial v} =
  e^{2\gamma -2\psi}. \eqno (A10)$$
This is, in fact, the Hamilton-Jacobi equation for ``radial" 
timelike geodesics, and ${\cal S}$ along the path differs from 
the proper time by a constant.

It is simple to illustrate a class of solutions of the eikonal 
equation
$(A10)$ via separation of variables, i.e.

$${\cal S}(u, v) = -c_{1} \alpha^{p} (u) - c_{2} \beta^{q}(v), \eqno (A11)$$
where $p = 4 -\sqrt{6},\  q=4+\sqrt{6},$ and $c_{1}$ and $c_{2}$ are constants such that

$$c_{1} c_{2} = {1 \over 5 l^{2}}. \eqno (A12)$$
Equations $(A8)$ and $(A9)$ now have solutions

$$\alpha(u) = \left ({\tau_{1} - \tau \over 2c_{1}}\right ) ^{1/ p}, 
\eqno (A13)$$
$$\beta(v) = \left ( {\tau_{2} - \tau \over 2c_{2}}\right )^{1 / q}, 
\eqno (A14)$$
where $\tau_{1}$ and $\tau_{2}$ are constants of integration. 
More explicitly, let us consider the normal form of the metric with 
$\alpha (u) = l_0^{-1/2} u$ and $\beta (v) = l_0^{-1/2} v$ as employed in
section 4. The geodesic would then hit the null singular 
hypersurface $u = 0$ (or $v = 0)$ at $\tau = \tau_{1}$ 
(or $ \tau = \tau_{2})$.
The singular nature of these moving cylinders can be seen from 
the fact that the geodesic can not be continued past them since then $u$ or $v$ would become complex.
\vskip1cm
{\bf
\noindent APPENDIX B: THE LANDAU-LIFSHITZ PSEUDOTENSOR IN RIEMANN NORMAL 
COORDINATES
}
\vskip 15pt
The energy of a gravitational field --- if it can be defined at all --- is 
nonlocal according to general relativity. On the other hand, 
the Bel-Debever-Robinson 
tensor is locally defined. A connection could perhaps be established between 
these concepts if the energy-momentum pseudotensor of the gravitational 
field is expressed in Riemann normal coordinates about a typical event in 
spacetime.

Let $x^{\mu}$ be the Riemann normal coordinates in the neighborhood of some 
point (``origin") in spacetime; then,

$$g_{\mu\nu} = \eta_{\mu\nu} - {1 \over 3} R_{\mu\alpha\nu\beta} \ x^{\alpha} x^{\beta} + \cdots , \eqno (B1)$$

$$\Gamma^{\mu}_{\nu\rho} =- {1 \over 3} (R^{\mu}_{\ \nu\rho\sigma} +
  R^{\mu}_{\ \rho\nu\sigma}) x^{\sigma} + \cdots . \eqno (B2)$$
The Landau-Lifshitz pseudotensor is quadratic in the connection coefficients
by construction; therefore, $t^{L-L}_{\mu\nu}$ is --- at the lowest order ---
quadratic in Riemann normal coordinates. Hence,

$$t^{L-L}_{\mu\nu,\alpha\beta} = 
{c^{4} \over 144 \pi G} \Theta_{\mu\nu\alpha\beta}
  + \cdots , \eqno (B3)$$
where $\Theta_{\mu\nu\alpha\beta}$ is symmetric in its first and second 
pairs of indices by construction and is given by

$$\Theta_{\mu\nu\alpha\beta} = 
{1 \over 2} (R^{\rho\sigma}_{\ \ \mu\alpha}  R_{\nu\sigma\rho\beta} + 
R^{\rho\sigma}_{\ \ \mu\beta} R_{\nu\sigma\rho\alpha})
+ {7 \over 2} (R_{\mu\rho\sigma\alpha} R_{\nu\ \ \beta}^{\ \rho\sigma} +
  R_{\mu\rho\sigma\beta} R_{\nu\ \ \alpha}^{\ \rho\sigma} )$$
$$-{3 \over 8} \eta_{\mu\nu} \eta_{\alpha\beta} 
R_{\rho\sigma\kappa\delta} R^{\rho\sigma\kappa\delta}. \eqno (B4)$$
This expression should be compared and contrasted with equation $(57)$
that expresses the Bel-Debever-Robinson tensor in a similar form. 
There is no simple relationship between 
$\Theta_{\mu\nu\alpha\beta}$ and $T_{\mu\nu\alpha\beta}$; 
however, one can show that

$$\Theta_{\mu\nu\alpha\beta} -7 \ T_{\mu\nu\alpha\beta} = {1 \over 16} 
\eta_{\alpha\beta} \eta_{\mu\nu} K + {1 \over 4} 
(R^{\rho\sigma}_{\ \ \mu\alpha}R_{\rho\sigma\nu\beta} + 
 R^{\rho\sigma}_{\ \ \mu\beta} R_{\rho\sigma\nu\alpha}) 
\eqno (B5)$$
in Riemann normal coordinates. Here $K$ is the Kretschmann scalar, i.e. $K = 
R_{\mu\nu\rho\sigma} R^{\mu\nu\rho\sigma}$.
\vskip1cm
\noindent
{\bf REFERENCES}
\vskip.5cm
\item{[1]}  Mashhoon B, McClune J C and Quevedo H ``Gravitational
Superenergy Tensor'' pre\-print 1996

\noindent
\item{[2]} Wheeler J A 1977 {\it Phys. Rev. D} {\bf 16} 3384; Ib\'a\~nez J and Verdaguer E 1985
{\it Phys. Rev. D} {\bf 31} 251; Isenberg J, Jackson M and 
Moncrief V 1990 {\it J. Math. Phys.} {\bf 31} 517;  Br\'eton N, 
Feinstein A and Ib\'a\~nez J 1993 {\it Gen. Rel. Grav.} {\bf 25} 267

\noindent
\item{[3]} Mashhoon B  and  Quevedo H 1990 {\it Phys. Lett. A} {\bf 151} 464

\noindent
\item{[4]} Quevedo H and Mashhoon B 1991 in {\it Relativity and Gravitation:
Classical and Quantum}, Proceedings of the Conference SILARG VII, edited
by J. C. D'Olivo {\it et al.} (World Scientific, Singapore, 1991) 
pp. 287--293. There is an error on p. 289 of this paper in the discussion
of the exterior field of a rotating cylinder. Contrary to the statement in
the paper, the exterior vacuum field is not always static. It is, in fact,
stationary for high angular momentum. This has been shown by Bonnor W B
1980 {\it J. Phys. A} {\bf 13} 2121. B. M. is grateful to Professor Bonnor for
clarifying remarks. Moreover, a typographical error occurs in equation $(9):
\mu,_{uv}$ should be replaced by $\mu,_{vv}$.

\noindent
\item{[5]}  Ardavan H 1984 {\it Phys. Rev. D} {\bf 29} 207; 1989
{\it Proc. Roy. Soc. London A} {\bf 424} 113

\noindent
\item{[6]} Ardavan H 1984 in {\it Classical General Relativity}, edited by
Bonnor W B, Islam J N and MacCallum M A H (Cambridge University
Press, Cambridge, 1984) pp.5--14

\noindent
\item{[7]} Zakharov V D {\it Gravitational Waves in Einstein's Theory}
(Halsted, New York, 1973), translated by R N Sen from the 1972
Russian edition

\noindent
\item{[8]} Beck G 1925 {\it Z. Physik} {\bf 33} 713

\noindent
\item{[9]} Einstein A and Rosen N 1937 {\it J. Franklin Inst.} {\bf 223} 43;
Rosen N 1954 {\it Bull. Res. Council Israel} {\bf 3} 328

\noindent
\item{[10]} Bonnor W B 1957 {\it J. Math. Mech.} {\bf 6} 203;  Weber J
and Wheeler J A  1957 {\it Rev. Mod. Phys.} {\bf 29} 509

\noindent
\item{[11]} Van den Bergh N and Wils P 1986 
{\it Class. Quantum Grav.} {\bf 2} 229

\noindent
\item{[12]} Bel L 1958 {\it C. R. Acad. Sci. Paris} {\bf 247} 1094; 1959 {\it ibid.} {\bf 248}
1297; 1962 {\it Cah. Phys.} {\bf 16} 59; Debever R 1958 {\it Bull. Soc. Math. 
Belg.} {\bf 10} 112; Bel L 1962 in {\it Colloques Internationaux
du CNRS}, Paris, pp.119--126

\vfill\eject
\noindent
\centerline{\bf
FIGURE CAPTIONS
}
\vskip.5cm\noindent
Figure 1. The plot of $X(r)$ versus $r$ for three solutions of equation
$(40)$ with the boundary conditions that at $r=1,\ X=2$ and $X_{r} =
-1, 0, 1$. The function $-X(r)$ is plotted in this figure as well. Note that the solution for $X_{r}(r=1)=-1$ approaches the special
solution $(41)$, also depicted here, for $r\to 0$. The behavior of the
solution for $X_{r}(r=1)=1$ near $r_{0}\simeq 0.22$ is in accordance with
equation $(43)$.

\vskip1cm\noindent
Figure 2. The functions $\rho(\tau)/l_{0}$ and $\phi (\tau)$ given by 
equations $(59)$ and $(60)$, respectively, with $\phi_{0}=\pi/2$ are plotted
here as polar coordinates (i.e., abscissa $= \rho \ $cos$\phi$ and ordinate
$= \rho \ $sin$\phi$). This represents the trajectory of a particle following 
the ``radial'' geodesic given by equations $(58)-(61)$. The particle starts
at $\tau = 0$ from $\rho = 0$ and moves counterclockwise until it returns to
a singular axis $(\rho = 0)$ at $\tau = l_{0}$. The return trip takes only
about ten percent of the total proper time $l_{0}$.
\end